# Access Pattern-Based Code Compression for Memory-Constrained Embedded Systems*

O. Ozturk, H. Saputra, and M. Kandemir
Dept. of Computer Science and Engineering
The Pennsylvania State University
University Park, PA 16802, USA
{ozturk,saputra,kandemir}@cse.psu.edu

I. Kolcu
Computation Department
UMIST
Manchester M60 1QD, UK
ikolcu@umist.ac.uk

## Abstract

As compared to a large spectrum of performance optimizations, relatively little effort has been dedicated to optimize other aspects of embedded applications such as memory space requirements, power, real-time predictability, and reliability. In particular, many modern embedded systems operate under tight memory space constraints. One way of satisfying these constraints is to compress executable code and data as much as possible. While research on code compression have studied efficient hardware and software based code strategies, many of these techniques do not take application behavior into account, that is, the same compression/decompression strategy is used irrespective of the application being optimized. This paper presents a code compression strategy based on control flow graph (CFG) representation of the embedded program. The idea is to start with a memory image wherein all basic blocks are compressed, and decompress only the blocks that are predicted to be needed in the near future. When the current access to a basic block is over, our approach also decides the point at which the block could be compressed. We propose several compression and decompression strategies that try to reduce memory requirements without excessively increasing the original instruction cycle counts.

## 1. Introduction

Most embedded systems have tight bounds on memory space. As a consequence, the application designer needs to be careful in limiting the memory space demand of code and data. However, this is not a trivial task, especially for large-scale embedded applications with complex control structures and data access patterns. One potential solution to the memory space problem is to use data and code compression.

Prior research in code compression studied both static and dynamic compressions techniques, focusing in particular on efficient compression/decompression strategies [1, 3, 4, 5, 7, 17, 18, 19]. One potential problem with most of these techniques is that the compression and decompression decisions are taken in an application-insensitive manner; that is, the same compression/decompression strategy is employed for all applications independent of their specific instruction access patterns.

In this paper, we propose a control flow graph (CFG) centric approach to reducing the memory space consumption of executable binaries. The main idea behind this approach is to keep basic blocks of the application in the compressed form as much as possible, without increasing the original execution cycle counts excessively. An important advantage of doing so is that the executable code occupies less memory space at a given time, and the saved space can be used by some other (concurrently executing) applications. [1] The proposed approach achieves this by tracking the basic block accesses (also called the instruction access pattern) at runtime, and by invoking compressions/decompressions based on the order in which the basic blocks are visited. On the one hand, we try to save as much memory space as possible. On the other hand, we do not want to degrade the performance of the application significantly by performing frequent compressions and decompressions, which could potentially occur in the critical path during execution. This paper makes the following major contributions:

---

[1] Alternately, in embedded systems that execute a single application, the memory space saved can enable the use of a smaller memory, thereby impacting both form factor and overall cost. As a third option, saved memory space can be used to increase energy savings in banked memory architectures.

* This work was supported in part by NSF Career Award #0093082.

- It proposes a basic block compression strategy called the *k-edge algorithm* that can be used for compressing basic blocks whose current executions are over.
- It proposes a set of basic block *pre-decompression* strategies, wherein a basic block is decompressed before it is actually needed, in an attempt to reduce the potential performance penalty that could be imposed by the on-line decompression.

The rest of this paper is organized as follows. Section 2 summarizes basic concepts regarding the control flow graph based code representation, and the assumptions we made about our execution environment. Sections 3 and 4 discuss the basic block compression and decompression strategies, respectively, proposed in this paper. Section 5 gives the details of our implementation. Section 6 discusses related work on code compression. Section 7 concludes the paper by summarizing our contributions.

## 2. Program Representation and Architecture

A control flow graph (CFG) is an abstract data structure used in compilers to represent a procedure/subprogram [20]. Each node in the CFG represents a *basic block*, i.e., a straight-line piece of code without any jumps or jump targets; jump targets start a block, and jumps end a block. In this graph, jumps in the control flow are represented by directed edges. There are two specially designated blocks: the *entry block*, through which control enters into the flow graph, and the *exit block*, through which all control flow leaves. The CFG is essential to several compiler optimizations based on global dataflow analysis such as def-use chaining and use-def chaining [20]. It should be emphasized that a CFG is a static (and conservative) representation of the program, and represents all the alternatives of control flow (i.e., all potential execution paths). As an example, both arms of an if-statement are represented in the CFG, while in a specific execution (with a particular input), only one of them could actually be taken. A cycle in the CFG may imply that there is a loop in the application code. Figure 1 depicts an example CFG fragment that contains two loops.

The approach proposed in this paper saves memory space by compressing basic blocks as much as possible without unduly degrading performance. We assume a software-controlled code memory either in the form of an external DRAM or in the form of an on-chip SRAM (e.g., a scratch-pad memory [21]). It must be emphasized that our main objective in this study is to reduce the memory space requirements of embedded applications. However, if there is another level of memory in front of the memory where our approach targets (i.e., a memory between the target memory and the CPU), the proposed approach also brings reductions in memory access latency (as we need to read less amount of data from the target memory) as well as in the energy consumed in bus/memory accesses. However, a detailed study of these issues is beyond the scope of this paper.

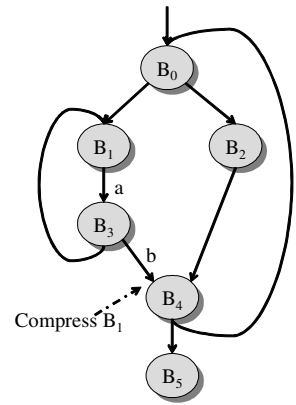

**Figure 1. An example CFG fragment. Assuming that the execution takes the left branch following $B_0$, the 2-edge algorithm (i.e., the k-edge algorithm with k=2) starts compressing $B_1$ just before the execution enters basic block $B_4$**

Another important issue is that, while in most of the cases discussed in this paper we do not put a restriction on the total memory space that could be used by the application being optimized, our approach needs only a slight modification to address this issue. Specifically, all that needs to be done is to check before each basic block decompression whether this decompression could result in exceeding the maximum allowable memory space consumption, and if so, compress one of the decompressed basic blocks that are in the uncompressed form. One could use LRU or a similar strategy to select the victim basic block when necessary.

## 3. Basic Block Compression

In this section, we discuss the *k-edge algorithm* in detail. This algorithm compresses a basic block that has been visited by the execution thread when the $k^{th}$ edge following its visit is traversed. It is to be noted that the k parameter can be used to tune the aggressiveness of compression. Consequently, the k-edge algorithm actually specifies a family of algorithms (e.g., 1-edge, 2-edge, 10-edge, etc). For example, let us consider the CFG illustrated in Figure 1. Assuming that we have visited basic block $B_1$ and, following this, the execution has traversed the edges marked as a and b, the 2-edge algorithm (i.e., the k-edge algorithm with k=2) starts compressing $B_1$ just before the execution enters basic block $B_4$.

Selecting a suitable value for the k parameter is important as it determines the tradeoff between memory space saving and performance overhead. Specifically, if we use a very small k value, we aggressively compress basic blocks but this may incur a large performance penalty for the blocks with high temporal reuse (though it is beneficial from a memory space viewpoint). In other words, if a basic block is revisited within a short period of time, a small k value could entail frequent compressions and decompressions (note that a basic block can be executed only when it is not in the compressed form). On the other hand, a very large k value delays the compression, which may be preferable from the performance angle (as it increases the chances of finding a basic block in the uncompressed form during execution when it is reached). But, it also increases the memory space consumption.

Another important issue is how one can perform compressions. Note that, in a single-threaded execution, the compression comes in the critical path of execution, and can slow down the overall execution dramatically. Therefore, we propose a *multi-threaded* approach, wherein there exists a *compression thread* (in addition to the main *execution thread*), whose sole job is to compress basic blocks at the background, thereby incurring minimal impact on performance. Specifically, the compression thread utilizes the idle cycles of the execution thread to perform compressions.

## 4. Basic Block Decompression

We have at least two options for performing basic block decompressions. In the first option, called the *on-demand decompression* (also called the *lazy decompression*), a basic block is decompressed only when the execution thread reaches it. That is, basic block decompressions are performed on a need basis. The most important advantage of this strategy is that it is easy to implement since we do not need an extra thread to implement it. All we need is a bit per basic block to keep track of whether the block accessed is currently in the compressed form or not. Its main drawback is that the decompressions can occur in the critical path, and thus degrade performance significantly. In the second option, referred to as the *pre-decompression*, a basic block is decompressed before it is actually accessed. The rationale behind this approach is to eliminate (or, at least reduce) the potential delay that would be incurred as a result of decompression. In other words, by pre-decompressing a basic block, we are increasing the chances that the execution thread finds the block in the uncompressed form, thereby not losing any extra execution cycles for decompressing it. This pre-decompression based scheme has, however, two main problems. First, we need a *decompression thread* to implement it. Second, pre-decompressing a basic block ahead of time can increase the memory space consumption.

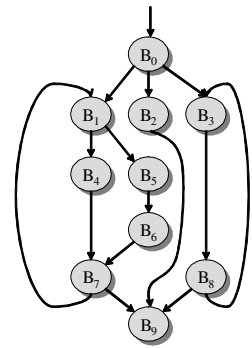

**Figure 2. An example CFG fragment that can be optimized using pre-decompression**

It is easy to see that a pre-decompression based scheme can be implemented in different ways. In this paper, we study this issue by focusing on two different dimensions. First, we have a choice in selecting the basic block(s) to pre-decompress. Second, we have a choice in selecting the time to pre-decompress them.[2] These two choices obviously bring associated performance/memory space tradeoffs. For example, pre-decompressing more basic blocks increases the chances that the next block to be visited will be in the uncompressed form (which is preferable from the performance viewpoint provided that we are able to hide the decompression cost); but, it also increases the memory space consumption. Similarly, pre-decompressing basic blocks early (as compared to pre-decompressing them at the last moment) involves a similar tradeoff between performance and memory space consumption.

In this paper, we explore this two-dimensional pre-decompression search space using two techniques. First, to determine the point at which we initiate decompression, we use an algorithm similar to k-edge. In this algorithm (also called *k-edge*), a basic block is decompressed (if it is not already in the uncompressed form) when there are *at most* k edges that need to be traversed before it could be reached. As before, k is a parameter whose value can be tuned for the desired memory space – performance overhead tradeoff. An example is depicted in Figure 2. Assuming k=3, in this figure, basic block $B_7$ is decompressed at the end of basic

---

[2] At this point, the analogyy between pre-decompression and software-initiated data/code pre-fetching should be noted. The two choices mentioned in the text correspond to selecting the blocks to prefetch and timing of prefetch in the context of prefetching.

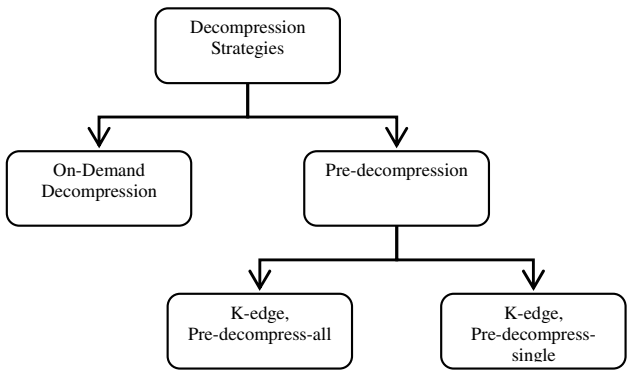

**Figure 3. The decomposition design space explored in this work. For compression, we always use the k-edge algorithm**

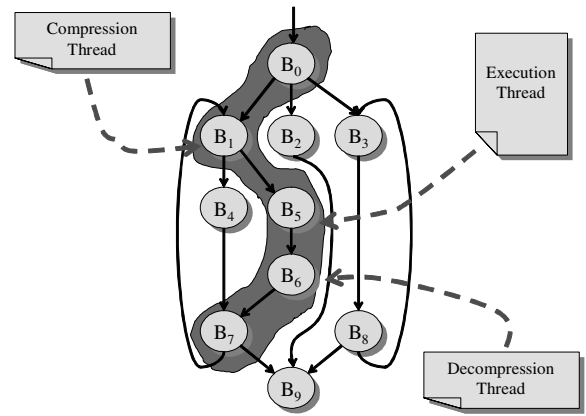

**Figure 4. The cooperation between the three threads during execution. The execution thread follows the decompression thread, and the compression thread follows the execution thread**

block $B_1$ (i.e., when the execution thread exits basic block $B_1$, the decompression thread starts decompressing $B_7$). This is because, from the end of $B_1$ to the beginning of $B_7$, there are at most 3 edges that need to be traversed. Second, to determine the basic block(s) to decompress, we use a *prediction-based* strategy. The idea is to determine the basic block that could be accessed next and to pre-decompress it ahead of time. In this paper, we discuss two different prediction-based strategies. In the first strategy, called *pre-decompress-all*, we pre-decompress all basic blocks that are at most k edges away from the exit of the currently processed block. In the second strategy, called *pre-decompress-single*, we select only one basic block among all blocks that are at most k edges ahead of the currently processed basic block. It is to be noted that while pre-decompress-all favors performance over memory space consumption, pre-decompress-single favors memory space consumption over performance. To demonstrate the difference between these two pre-decompression based strategies, we consider the CFG fragment in Figure 2 again, assuming this time, for illustration purposes, that blocks $B_4$, $B_5$, $B_8$, and $B_9$ are currently in the compressed form, all other blocks are in the uncompressed form, and the execution thread has just left basic block $B_0$. Assuming further that k=2, in the pre-decompress-all strategy, the decompression thread decompresses $B_4$, $B_5$, $B_8$, and $B_9$. In contrast, in the pre-decompress-single strategy, we predict the block (among these four) that is to be the most likely one to be reached than the others, and decompress only that block. Figure 3 shows the decompression design space explored in this paper.

Figure 4 summarizes our approach to code compression for reducing memory space consumption. It is assumed that the highlighted path is the one that is currently taken by the execution thread. In the ideal case, the decompression thread traverses the path before the execution thread and decompresses the basic blocks on it so that the execution thread finds them directly in the executable state. The compression thread, on the other hand, follows the execution thread and compresses back the basic blocks whose executions are over. The k parameters control the distance between the threads.

## 5. Implementation Details

In implementing the compression/decompression-based strategy described, there is an important challenge that needs to be addressed. Specifically, when a basic block is compressed or decompressed, the branch instructions that target that block must be updated. In addition, the saved memory space (as a result of compressions) should be made available to the use of other applications with minimum overhead. In particular, one may not want to create too much memory fragmentation. This is because an excessively fragmented free space either cannot be used for allocating large objects or requires memory compaction to do so. Therefore, our current implementation slightly deviates from the discussion so far, in particular when compressions are concerned. Specifically, we start with a memory image, wherein all basic blocks are stored in their compressed form. Note that this is the minimum memory that is required to store the application code. As the execution progresses, we decompress basic blocks (depending on the instruction access pattern and the decompression strategy adopted, as discussed earlier), and store the decompressed (versions of the) blocks in a separate location (and keep the compressed versions as they are). Later, when we want to compress the block, all we need to do is to delete the decompressed version. In this way, the compression process does not take too much time. In addition, the memory space is not fragmented too much as the locations of the compressed blocks do not change during execution. We illustrate the idea using the example in Figure 5 with on-demand decompression. The figure shows an example CFG fragment, and traces the sequence of events for a particular execution scenario. Initially, all the basic blocks are in the compressed form and stored in the compressed code area. The program counter (PC) points to the entry of the first basic block, which is $B_0$ in this case (1). Fetching an instruction from the compressed code area triggers a memory protection exception.

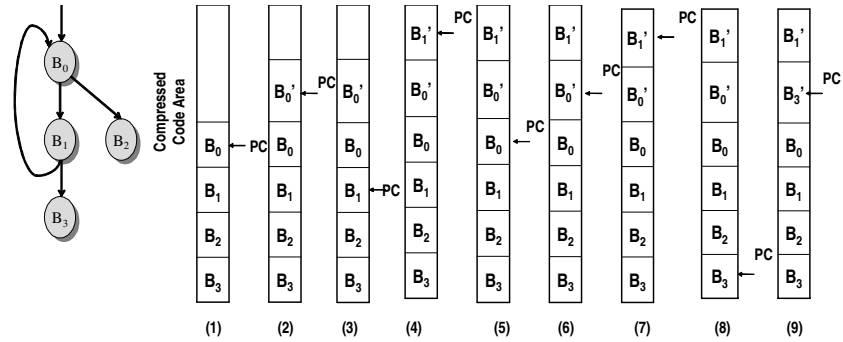

**Figure 5. An example CFG fragment and the contents of the instruction memory when the basic block access pattern is $B_0$, $B_1$, $B_0$, $B_1$, and $B_3$**

The exception handler decompresses block $B_0$ into $B_0'$ and sets PC to the entry of $B_0'$ (2). Assuming that block $B_1$ is the one that follows $B_0$, after the execution of block $B_0$, the PC points to the entry of block $B_1$ (3). Since $B_1$ is in the compressed code area, the exception handler is invoked to decompress $B_1$ into $B_1'$ and update the target address of the branch instruction in $B_0$ and set the PC to the entry of $B_1'$ (4). Let us now assume that the execution thread next visits $B_0$ again. Consequently, after the execution of $B_1'$, we branch to the entry of $B_0$ (5). At this time, we do not need to decompress $B_0$ once again. The exception handler updates the target address of the last branch instruction of block $B_1'$ to the entry of $B_0'$, and subsequently sets PC to the entry of $B_0'$ (6). Following $B_0'$, the execution thread can branch to $B_1'$ directly without generating any exception (7). Let us assume now that the execution next visits $B_3$. Consequently, the PC points to the entry of this basic block (8). Assuming that our compression strategy uses k=2, at this point, we delete the decompressed version of $B_0$ (which is $B_0'$), and decompress $B_3$ into $B_3'$ as illustrated in (9). It is to be noted that, when we discard a decompressed block, we also need to update the target addresses of the branch instructions (if any) that jump to the discarded block. For this purpose, for each decompressed block, we also maintain a "remember set" that records the addresses of the branch instructions that jump to this block.

Another issue is how to keep track of the fact that k edges have been traversed so that we can delete the decompressed version of the blocks. Our current implementation works as follows. For each basic block, we maintain a counter, which is reset to zero when the basic block is executed. At each branch, the counter of each (uncompressed) basic block is increased by 1 and (the decompressed versions of) the basic blocks whose counter reaches k are deleted.

## 6. Discussion of Related Work

Many embedded systems rely on special hardware to execute compressed code, such as Thumb for ARM processors [2], CodePack [14] for PowerPC processors, and MIPS16 [13] for MIPS processors. However, the requirement for special hardware limits their general applicability. Lefurgy et al [16] propose a hybrid approach that decompresses the compressed code at the granularity individual cache lines. Kirovski et al [15] present a procedure-based compression strategy that requires little or no hardware support. There has been a significant amount of work that explores the compressibility of program representations [12]. The resulting compressed form either must be decompressed (or compiled) before execution [7, 11, 8], or it can be executed without decompression [5, 9]. A hybrid approach is to use an interpreter to execute the compressed code [10, 22]. Compared to the direct execution approach, the interpreter-based approach usually allows more complex coding schemes, and thus, achieves smaller memory consumption for the compressed code. However, the interpreter itself occupies memory space.

Debray and Evans [6] present a code compression strategy that operates at a function granularity, i.e.,

functions constitute compressible units. Their work exploits the property that for most programs, a large fraction of the code is rarely touched. Our work is different from theirs in at least two aspects. First, we operate on a finer granularity (basic block level). Therefore, we can potentially save more memory space (when, for example, a particular basic block chain within a large function is repeatedly executed, in which case our approach can keep the unused memory blocks—in the function—in the compressed form). Second, we also employ pre-decompression that helps us reduce the negative impact of compression on performance.

## 7. Summary

This paper proposed a novel code compression strategy that is guided by the control flow graph (CFG) representation of an embedded program. In this strategy, the unit of compression/decompression is a single basic block of code. Conceptually, our approach employs three threads: one for compressing basic blocks, one for decompressing them, and one for executing the application code. We presented several pre-decompression techniques wherein a basic block is decompressed before it is actually needed, in an attempt to reduce the potential performance penalty caused by decompression

## References

[1] B. Abali et al. "Memory Expansion Technology (MXT): Software Support and Performance." IBM Journal of Research and Development, Vol 45, No 2, 2001.

[2] "An Introduction to Thumb," Adanced RISC Machines Ltd., March 1996. http://www.win.tue.nl/cs/ps/rikvdw/papers/ARM95.pdf

[3] L. Benini, D. Bruni, A. Macii, and E. Macii. "Hardware-Assisted Data Compression for Energy Minimization in Systems with Embedded Processors." In Proc. Design, Automation and Test in Europe Conference & Exhibition, 2002.

[4] L. Benini, A. Macii, E. Macii, and M. Poncino. "Selective Instruction Compression for Memory Energy Reduction in Embedded Systems." In Proc. International Symposium on Low Power Electronics and Design , 1999. pp 206 - 211.

[5] K. Cooper and N. McIntosh. "Enhanced Code Compression for Embedded RISC Processors," In Proc. ACM Conference on Programming Language Design and Implementation, pp. 139-149, May 1999.

[6] S. Debray and W. S. Evans. "Cold Code Decompression at Runtime," Communications of the ACM, pp. 54-60, 2003.

[7] J. Ernst, W. Evans, C. W. Fraser, T. A. Proebsting, and S. Lucco. "Code Compression," In Proc. ACM Conference on Programming Language Design and Implementation, pp. 358-365, Las Vegas, NV, 1997.

[8] M. Franz and T. Kistler. "Slim Binaries," Communication of the ACM 40, 12 (Dec), 1997, pp. 87-94.

[9] C. Fraser, E. Myers, and A. Wendt. "Analyzing and Compressing Assembly Code," In Proc. Symposium on Compiler Construction, Volume 19, pages 117-121, June 1984.

[10] C. W. Fraser and T. A. Proebsting. "Custom Instruction Set for Code Compression," Unpublished manuscript. http://research.microsoft.com/~toddpro/papers/pldi2.ps, October 1995.

[11] [M. Franz. "Adaptive Compression for Syntax Trees and Iterative Dynamic Code Optimization: Two Basic Technologies for Mobile-Object Systems," In J. Vitek and C. Tschudin, editors, Mobile Object Systems: Towards the Programmable Internet, LNCS vol. 1222, pp. 263-276. Springer, February 1997.

[12] J. Hoogergrugge, L. Augusteijn, J. Trum, and R. Van De Wiel. "A Code Compression System Based on Pipelined Interpreters," Software Practice and Experience, 29(1), pp. 1005-1023, January 1999.

[13] K. D. Kissel. "MIPS16: High-Density MIPS for the Embedded Market," In Proc. Real Time Systems'97, 1997.

[14] T. M. Kemp et al. "A Decompression Core for PowerPC," IBM Journal of Research and Development, Vol. 42, No. 6, November 1998.

[15] D. Kirovski, J. Kin, and W. Mangione-Smith. "Procedure-Based Program Compression," In Proc. International Symposium on Microarchitecture, pp. 204-213, 1997.

[16] C. Lefurgy, E. Piccininni, and T. Mudge. "Reducing Code Size with Run-Time Decompression," In Proc. the Sixth International Symposium on High-Performance Computer Architecture, January 2000, pp. 218-227.

[17] C. Lefurgy, E. Piccininni, and T. Mudge. "Evaluation of a High-Performance Code Compression Method," In Proc. the 32nd Annual ACM/IEEE International Symposium on Microarchitecture, pp. 93-102, Haifa, Israel, 1999.

[18] H. Lekatsas, J. Henkel, and W. Wolf. "Code Compression for Low-Power Embedded System Design," In Proc. Design Automation Conference, 2000.

[19] H. Lekatsas, J. Henkel, and W. Wolf. "Code Compression as a Variable in Hardware/Software Codesign," In Proc. the Eighth International Workshop on Hardware/Software Codesign, 2000.

[20] S. Muchnick. "Advanced Compiler Design and Implementation." Morgan-Kaufmann, 1997.

[21] P. R. Panda, N. Dutt, and A. Nicolau. "Memory Issues in Embedded Systems-On-Chip: Optimizations and Exploration." Kluwer Academic Publishers, Norwell, MA, 1999.

[22] T.A. Proebsting. "Optimizing an ANSI C Interpreter with Superoperators," In Proc. Symposium on Principles of Programming Languages, pp 322-332, January 1995.